\documentclass[conference]{IEEEtran}

\IEEEoverridecommandlockouts

\usepackage{cite}
\usepackage{amsmath,amssymb,amsfonts}
\usepackage{relsize}
\usepackage{algorithmic}
\usepackage{graphicx}
\usepackage{textcomp}
\usepackage{xcolor}
\usepackage{hyperref}

\newtheorem{theorem}{Theorem}
\newtheorem{lemma}{Lemma}

\begin{document}

\title{Wideband and Entropy-Aware Deep Soft Bit Quantization
\thanks{This work was supported by NSF IFML 2019844, ONR grant N00014-19-1-2590, a UT Austin Machine Learning Lab Research Award, and an AWS Machine Learning Research Award.}
}

\author{\IEEEauthorblockN{Marius Arvinte}
\IEEEauthorblockA{\textit{Electrical and Computer Engineering} \\
\textit{University of Texas at Austin}\\
Austin, TX, USA \\
arvinte@utexas.edu}
\and
\IEEEauthorblockN{Jonathan I. Tamir}
\IEEEauthorblockA{\textit{Electrical and Computer Engineering} \\
\textit{University of Texas at Austin}\\
Austin, TX, USA \\
jtamir@utexas.edu}
}

\maketitle

\begin{abstract}
Deep learning has been recently applied to physical layer processing in digital communication systems in order to improve end-to-end performance. In this work, we introduce a novel deep learning solution for soft bit quantization across wideband channels. Our method is trained end-to-end with quantization- and entropy-aware augmentations to the loss function and is used at inference in conjunction with source coding to achieve near-optimal compression gains over wideband channels. To efficiently train our method, we prove and verify that a fixed feature space quantization scheme is sufficient for efficient learning. When tested on channel distributions never seen during training, the proposed method achieves a compression gain of up to $10 \%$ in the high SNR regime versus previous state-of-the-art methods. To encourage reproducible research, our implementation is publicly available at \href{https://github.com/utcsilab/wideband-llr-deep}{\texttt{https://github.com/utcsilab/wideband-llr-deep}}.
\end{abstract}


\section{Introduction}
Soft bit quantization \cite{novak2009quantization,rave2009quantization} is an important task in integrated, low-power digital communication platforms where memory is an expensive asset \cite{akeela2018software}. A critical application area for quantizing estimated soft bits is given by hybrid automatic repeat request (HARQ) schemes in, e.g., 5G networks \cite{anand2018resource}, where information from a failed transmission is stored in order to boost the performance via soft combining methods \cite{frenger2001performance}. Given that a cellular base station may communicate with hundreds of users simultaneously, storing soft bits from failed packets requires efficient and low-distortion quantization methods to avoid memory bottlenecks on the platform. Another application area where a flexible trade-off between compression rate and reconstruction distortion is desirable is given by compress-and-forward relaying schemes \cite{wu2013optimal,ghallab2018compress}, where estimated soft bits are forwarded to a receiver and compressed in order to lower relay channel resource utilization.

The recent success of deep learning applied to compression problems \cite{agustsson2017soft} motivates us to develop deep soft bit quantization methods. A major challenge here is given by the fact that any hard quantization operator has zero gradient almost everywhere, and thus cannot be used in conjunction with modern optimization algorithms. To this end, various types of practical approximations and solutions have been developed \cite{bengio2013estimating,agustsson2017soft,jung2019learning} to make deep neural networks \textit{quantization-aware}, and in this work, we use the pass-through gradient estimation approach \cite{bengio2013estimating} due to its simplicity and ease of implementation.

In this paper, we introduce a data-driven approach for soft bit quantization over wideband channels. Our scheme consists of a lightweight, properly initialized deep autoencoder network, and is trained for soft bit reconstruction in random channels using a differentiable approximation to quantization, as well as a continuous approximation to the entropy of a discrete source. During inference, lossless source coding is performed over the latent representations of soft bits from a \textit{wideband} channel transmission to maximize compression gains. Experimental results over simulated EPA \cite{instance1290} channel realizations demonstrate state-of-the-art performance of the proposed approach, as well as a controllable trade-off between compression rate and distortion.

\subsection{Related Work}
Prior work on soft bit quantization generally belongs in one of two categories: classical methods \cite{rave2009quantization,novak2009quantization,winkelbauer2015quantization} develop near-optimal scalar quantization methods directly in the log or hyperbolic tangent domain. In particular, the method in \cite{winkelbauer2015quantization} introduces an optimal scalar quantization method for soft bits that supports a data-driven formulation and learns a codebook that maximizes the mutual information between the original and reconstructed soft bits. While this is optimal for scalar (per soft bit position) quantization, it does not take advantage of redundancy in soft bits derived from the same or correlated channels.

Recently, the work in \cite{arvinte2019deep} introduces an architecture for deep soft bit quantization, building on the observation that the soft bits corresponding to a high-order modulation scheme transmitted over a single channel are correlated, and can always be represented exactly with three values, regardless of the modulation order. This motivates a deep learning approach in which an autoencoder is trained to compress the soft bits, which are further numerically quantized at inference time. Our work builds directly upon \cite{arvinte2019deep}, with the following important distinctions: (i) our approach is entropy- and quantization-aware during training, (ii) during inference, we apply source coding to compress soft bits, and (iii) provide a tunable, continuous trade-off between compression rate and distortion.

\subsection{Contributions}
Summarized, our contributions are the following:
\begin{enumerate}
    \item We introduce a deep soft bit quantization architecture that is quantization-aware through a differentiable approximation used in the backwards pass and entropy-aware through a soft entropy that is annealed over the course of training.
    \item We derive the exact variance for the latent representation of the soft bits in the asymptotically large signal-to-noise ratio (SNR) regime, at initialization, in a deep neural network with one hidden layer and $\textrm{ReLU}$ activation. This is used for the one-time design of a fixed quantization codebook and helps stabilize learning.
    \item We experimentally demonstrate state-of-the-art quantization performance and rate-distortion trade-off in realistic wideband channel models, when compared against classical and deep learning baselines, on a channel distribution that is completely unseen during training.
\end{enumerate}

\section{System Model}

\begin{figure*}[!t]
\centering
\includegraphics[width=1.\linewidth]{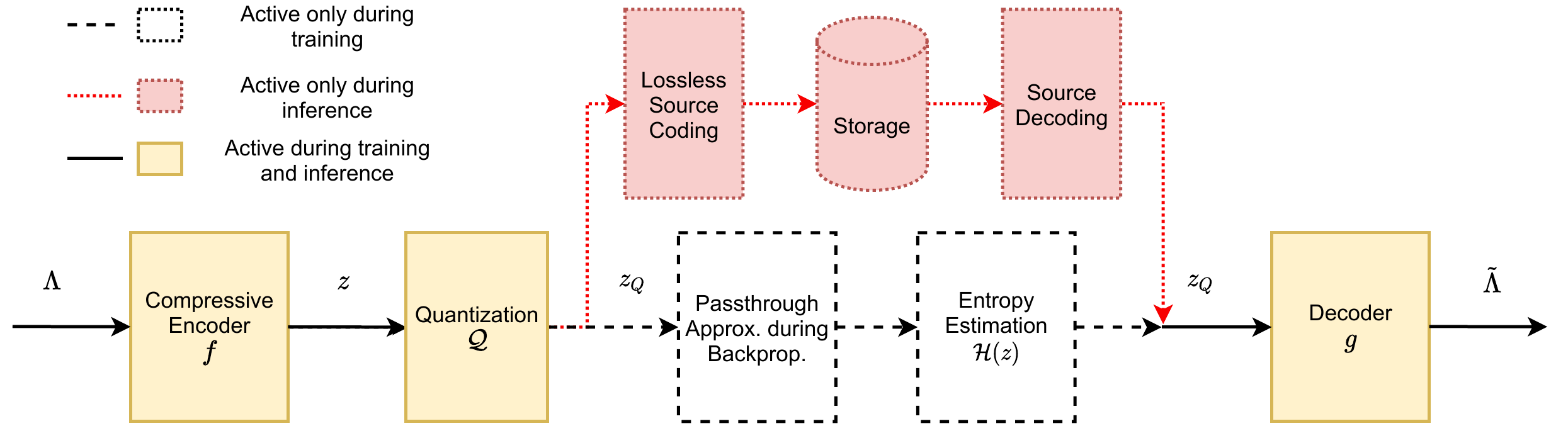}
\caption{Diagram of the proposed wideband quantization approach. Blocks in yellow are active during both training and inference (real-world deployment). The pass-through approximation and entropy estimation are only performed for training purposes. During inference, source coding is applied to the \textit{entire} wideband quantized latent matrix $\mathbf{z}_Q$, and the binary codeword is stored for future use at a potentially different location that has access to the decoder $g$.}
\label{fig:high_level}
\end{figure*}

We consider a communications model that transmits a number of $N$ parallel data channels to a single user, such as sub-carriers in an orthogonal frequency division multiplexing (OFDM) scenario. Assuming no cross-channel interference, the signal received on the $i$-th channel is given by the linear model \cite[Eq.~3.1]{tse2005fundamentals}:
\begin{equation}
    y_i = h_i x_i + n_i,
\end{equation}

\noindent where $h_i$ is the channel gain, $x_i$ is the transmitted symbol and $n_i$ is the noise corresponding to the $i$-th subcarrier, with all variables being complex-valued. We assume that the noise is drawn from a complex, circular Gaussian distribution with zero mean and standard deviation of $\sigma_n$. We assume that the symbols $x_i$ are obtained by mapping a set of $K$ bits $\{b_k\}_{k=1}^{K}$ to a complex-valued constellation symbol, which is the common practice of digital modulation. Given ideal channel knowledge $h_i$ and known noise statistics $\sigma_n$, and assuming that transmitted bits are sampled i.i.d. with equal probabilities, the maximum likelihood (ML) estimate of the log-likelihood ratio for the $k$-th bit transmitted on the $i$-th channel is given by:
\begin{equation}
    L_{i, k} = \log \frac{P(y_i | b_{i, k}=1)}{P(y_i | b_{i, k}=0)} = \log \frac{\mathlarger{\sum\limits_{s\in\{s | b_k = 1\}}} \exp{- \frac{|y_i - h_i s|^2}{\sigma_n^2} }}
    {\mathlarger{\sum\limits_{s\in\{s | b_k = 0\}}} \exp{- \frac{|y_i - h_i s|^2}{\sigma_n^2} }}.
    \label{eq:llr_def}
\end{equation}

The \textit{soft bits} are defined as $\Lambda_{i, k} = \tanh{\frac{L_{i, k}}{2}}$ and grouped in the \textit{wideband} soft bit matrix $\mathbf{\Lambda}$. The goal of wideband quantization is to design the triplet of functions $(f, g, \mathcal{Q})$, where the encoder $f$ maps the floating point input $\mathbf{\Lambda}$ to a latent representation, $\mathcal{Q}$ maps this representation to a finite bit string, and the decoder $g$ recovers $\mathbf{\Lambda}$ with minimal distortions. Note that this scheme maps the \textit{entire} soft bit matrix to a single binary codeword, and is thus a form of vector quantization. The compression rate and the distortion of the reconstruction are denoted by the functions:
\begin{equation}
    \mathcal{R} = H(f(\mathbf{\Lambda})) \ \text{and} \ \mathcal{D} = \frac{1}{N K} \sum_{i,k} \frac{|\Lambda_{i, k} - \tilde{\Lambda}_{i, k}|^2}{|\Lambda_{i, k}| + \epsilon},
\end{equation}
\noindent respectively, where $H$ represents the entropy of a discrete source expressed in bits, and $\tilde{\mathbf{\Lambda}} = g(f(\mathbf{\Lambda}))$ is the recovered soft bit matrix and $\epsilon$ is a small numerical constant used to prevent overflow. We use a sample-weighted version of the mean squared error, which is a pseudo-metric since it is not symmetric and does not satisfy the triangle inequality property. This choice is taken from \cite{arvinte2019deep}, since minimizing this metric places more importance on uncertain soft bits and benefits the decoding of error-correcting codes \cite{hemati2006dynamics}.

\section{Proposed Method}
Fig.~\ref{fig:high_level} shows an overview of the proposed approach for wideband, entropy-aware soft bit quantization. During training, the model uses the wideband soft bit matrix $\mathbf{\Lambda}$ and a soft entropy estimate of the quantized representation to optimize the weights of $f$ and $g$, which are deep neural networks. During inference, lossless source coding is applied to the quantized representation $\mathbf{z}_Q$ to reduce storage costs to near-entropy levels. The resulting binary string is stored until decoding is required, e.g., in hybrid ARQ or relay scenarios. In the following, we give a description of each of the involved components.

\subsection{Encoder}
The encoder $f$ is a function that maps an input matrix $\mathbf{\Lambda}$ to a latent matrix $\mathbf{z}$ by applying the same functional backbone in a row-wise manner and stacking the representations in a matrix:
\begin{equation}
    \mathbf{z} = \textrm{stack}_i(f(\mathbf{\Lambda}_i)).
\end{equation}

We design the backbone of $f$ as a fully-connected, feed-forward network with ReLU activation in the hidden layers and $\tanh{x}$ as the output activation. The input size is a vector of size $K$, the hidden layers are all of size $4K$, while the output is of fixed size equal to \textit{three}. This corresponds to the universal latent dimension of a soft bit vector with arbitrary $K$, as introduced in \cite{arvinte2019deep}. That is, without quantization, such a compressive representation (from $K$ soft bits to three latent variables) is guaranteed to exist and can be represented and learned by a deep neural network with a sufficient modeling capacity.

\subsection{Latent Quantization}
This block applies a discrete quantization operator $\mathcal{Q}$ to each component of the latent representation $\mathbf{z}$ in the forward pass of the network. During training, since the quantization operator has zero gradient almost everywhere, we use a pass-through approximation \cite{bengio2013estimating} to obtain a differentiable function for the backward pass. That is, the forward and backward pass signals are, respectively:
\begin{equation}
\begin{split}
z_{Q, i, \textrm{fw}} & = \mathcal{Q}(z_i), \\
z_{Q, i, \textrm{bw}} & = \texttt{sg}[\mathcal{Q}(z_i) - z_i] + z_i,
\end{split}
\end{equation}
\noindent where $\mathbf{z}_{Q, \textrm{fw}}$ and $\mathbf{z}_{Q, \textrm{bw}}$ are the forward and backward pass latent signals, respectively, and $\texttt{sg}$ is the stop-gradient operator, which prevents gradient from flowing in the backward pass. This leads to the gradient of the quantized representation with respect to its input being $\frac{\partial z_{Q,i,\textrm{bw}}}{\partial z_i} = 1$ and allows gradients to propagate to earlier layers.

Importantly, the function $\mathcal{Q}$ is a pre-determined scalar quantization function that is held fixed throughout learning and inference. This differentiates us from \cite{arvinte2019deep} and \cite{agustsson2017soft} and enables efficient learning, given a careful choice of $\mathcal{Q}$. In the following, we present a theoretical and empirical analysis of deep neural networks that are used for soft bit quantization in the high SNR regime and show that a choice for $\mathcal{Q}$ that avoids the issue of codebook collapse \cite{van2017neural,razavi2019generating} -- where a portion of the codebook may never be used during training --  can be found at initialization. We use the two following lemmas in our proof:
\begin{lemma}
Let $\mathbf{X} \in \mathbb{R}^{m \times n}$ be a matrix with i.i.d. Gaussian elements and let $\mathbf{Y} \in \mathbb{R}^{n}$ be an i.i.d. Rademacher random variable, independent of $\mathbf{X}$. Then, the elements of $\mathbf{X}\mathbf{Y}$ are distributed as i.i.d. Gaussian random variables.
\label{lemma:lemma_one}
\end{lemma}
\begin{IEEEproof}
Immediate by the independence of $\mathbf{X}$ and $\mathbf{Y}$. 
\end{IEEEproof}
\begin{lemma}
Let $X$ be a scalar random variable distributed as $\mathcal{N}(0, \sigma)$. Then, $\mathrm{relu}(X) = \mathrm{max}\{X, 0\}$ has the following properties:
\begin{itemize}
    \item $\mathbb{E}[\mathrm{relu}(X)] = \frac{1}{\sqrt{2\pi}} \sigma,$
    \item $\textrm{Var}(\mathrm{relu}(X)) = (\frac{1}{2} - \frac{1}{2\pi}) \sigma^2.$
\end{itemize}
\label{lemma:lemma_two}
\end{lemma}
\begin{IEEEproof}
Follows immediately from \cite[Page 3]{harva2007variational} and re-writing $\mathrm{relu}(X)$ as a mixture of two random variables.
\end{IEEEproof}
We now state and prove the following the following theorem.
\begin{theorem}
Let $f$ be a one hidden-layer, fully-connected neural network with no biases, hidden $\mathrm{relu}$ activation, and linear output activation. Let the length of the input vector $\mathbf{\Lambda}$ be $K$, the hidden size be $4K$, and the output size be $1$. The weight matrices are $\mathbf{W} \in \mathbb{R}^{4K \times K}$ and $\mathbf{V} \in \mathbb{R}^{1 \times 4K}$, respectively. We make the following assumptions:
\begin{itemize}
    \item The entries of $\mathbf{\Lambda}$ are drawn i.i.d. from a Rademacher distribution such that $p(\Lambda_i = 1) = p(\Lambda_i = -1) = 0.5$.
    \item The entries of the hidden layer weight matrix $\mathbf{W}$ are drawn i.i.d. from a Gaussian distribution with $\mu_w = 0$ and $\sigma_w = \sqrt{\frac{2}{5K}}$, respectively.
    \item The entries of the output layer weight vector $\mathbf{V}$ are drawn i.i.d. from a Gaussian distribution with $\mu_v = 0$ and $\sigma_v = \sqrt{\frac{2}{4K+1}}$, respectively.
\end{itemize}
Let $z = \sum_i v_i \mathrm{relu}(\mathbf{W}\mathbf{\Lambda})_i$ be the output of the network. Then, it satisfies the following properties:
\begin{itemize}
    \item $\mathbb{E}[z] = 0, $
    \item $\textrm{Var}(z) = \frac{8}{5} \frac{K}{4K+1}.$
\end{itemize}
\label{theorem:main}
\end{theorem}
\begin{IEEEproof}
Using Lemma~\ref{lemma:lemma_one} and the first two assumptions, it follows that the pre-activation values after the first layer are Gaussian distributed. Using Lemma~\ref{lemma:lemma_two} on these activations allows us to characterize the mean and standard deviation of $\mathrm{relu}(\mathbf{W}\mathbf{\Lambda})_i$. Since the entries $\mathbf{W}\mathbf{\Lambda}$ are i.i.d., it follows that the entries of $\mathrm{relu}(\mathbf{W}\mathbf{\Lambda})$ are also i.i.d., and also independent from the weights $\mathbf{V}$, as well as using the second and third assumptions. Since $\mathbf{V}$ is zero-mean, we obtain that:
\begin{equation}
\begin{split}
\textrm{Var}(z) & = \textrm{Var} \left( \sum_i v_i \mathrm{relu}(\mathbf{W}\mathbf{\Lambda})_i \right) \\
& = \sum_i \textrm{Var} \left( v_i \mathrm{relu}(\mathbf{W}\mathbf{\Lambda})_i \right) \\
& = \sum_i \textrm{Var}(v_i) \textrm{Var} \left( \mathrm{relu}(\mathbf{W}\mathbf{\Lambda})_i \right) \ + \\
& \ \ \ \textrm{Var}(v_i) \mathbb{E}\left[\mathrm{relu}(\mathbf{W}\mathbf{\Lambda})_i\right]^2 \\
& = 4K \sigma_v^2 (\sqrt{K} \sigma_w)^2 [(\frac{1}{2} - \frac{1}{2\pi}) + \frac{1}{2\pi}] \\
& = 2 K^2 \sigma_v^2 \sigma_w^2 \\
& = \frac{8}{5} \frac{K}{4K+1}.
\end{split}
\label{eq:theorem_proof}
\end{equation}
\end{IEEEproof}
\begin{figure}[!t]
\centering
\includegraphics[width=0.93\linewidth]{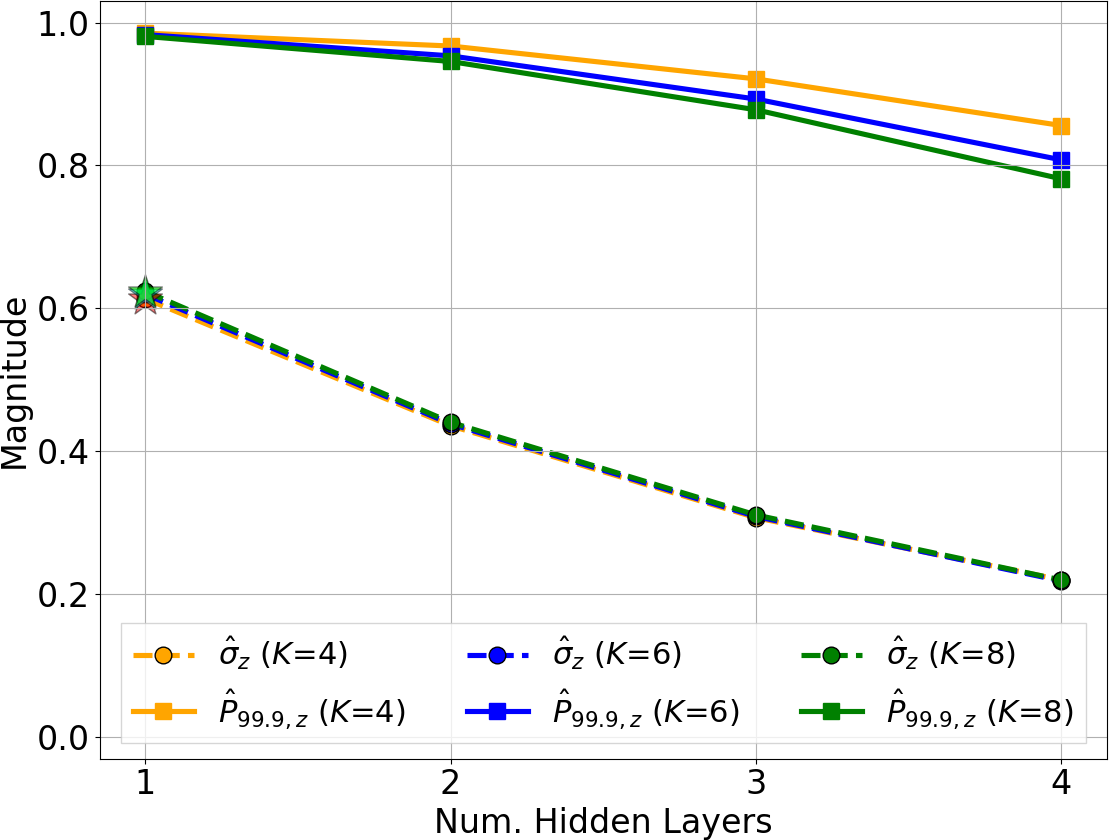}
\caption{Empirical verification of Theorem 1 across a varying number of $K$ values, corresponding to different modulation orders. The variables $\hat\sigma_z$ represent empirical estimates of the latent standard deviation \textit{at initialization}, while the stars mark the theoretical values (only available for one hidden layer). The variables $\hat{P}_{99.9, z}$ represent the 99.9 percentile values of the absolute latent variables at initialization.}
\label{fig:theory}
\end{figure}

The first assumption corresponds to operating in the asymptotically large SNR regime, where soft bits tend toward polarized values in the hyperbolic tangent domain. We study this regime since it allows exact analysis of the variance and serves as an upper bound for lower SNR regimes in terms of the latent space variance, since the distribution of the soft bits there is more biased toward zero and has a reduced variance.

The last two assumptions concern the deep neural network at initialization and match the Glorot weight initialization scheme \cite{glorot2010understanding}. As motivated in \cite{glorot2010understanding}, this initialization is carefully chosen such that the variance of the signal is reduced as the network gets deeper. To verify Theorem~\ref{theorem:main} and the empirical reduction of variance, we plot the estimated standard deviation of $\mathbf{z}$ and the estimated range of $\mathbf{z}$ in Fig.~\ref{fig:theory} for a varying depth of the network and different values of $K$. The architecture follows the exact assumptions of Theorem~\ref{theorem:main}, and is the basis for the model we use in practice.

The exact match between the empirical $\hat{\sigma}_z$ and the starred points verifies Theorem~\ref{theorem:main} for a network with one hidden layer, and these values are almost invariant to $K$ due to the ratio in (\ref{eq:theorem_proof}).  Fig.~\ref{fig:theory} also plots the $99.9$ percentile values for an increasing number of hidden layers. While an exact analysis is out of scope here, we find that using the Glorot initialization leads to a decreasing latent variance as the network gets deeper, as originally pointed out in \cite{glorot2010understanding}. Since the encoder $f$ uses a $\mathrm{tanh}$ function as activation, this is extremely useful in preventing latent collapse -- the presence of strong modes at $\pm 1$ -- and allows the use of a fixed quantizer $\mathcal{Q}$ throughout the entire training process.

\subsection{Entropy Estimation}
Given $\mathbf{z}$, $\mathbf{z}_Q$ and a quantization codebook $Q$ with $M$ entries, we estimate the $\tau$-soft entropy \cite{agustsson2017soft} as:
\begin{equation}
    \mathcal{H}(\mathbf{z}; \tau) = -\frac{1}{N} \sum_{i=1}^M \sum_{j=1}^N \phi_i(z_j; \tau) \log{p_i},
\end{equation}
\noindent where $q_{i, j} = \phi_i(z_j; \tau)$ represents the soft allocation of $z_j$ to the $i$-th entry in the quantization codebook. That is, $q_{i, j} = \textrm{softmax}_i\big(-\frac{|z_j - Q_i|^2}{\tau}\big)$, where the softmax is taken across all codebook entries and $\tau$ represents the \textit{inverse temperature} of this approximation. The terms $p_i$ represent the empirical probability estimates of $\mathbf{z}_Q$ obtained by counting over $N$ samples. Hence, no gradient flows through the $\log{p_i}$ term during training. An important aspect here is that as $\tau \rightarrow \infty$ and the sample size $N$ is sufficiently large, we have that $\mathcal{H(\mathbf{z};\tau)} \rightarrow H(\mathbf{z}_Q) = - \sum_i p_i \log{p_i}$, the entropy of the discrete random variable $\mathbf{z}_Q$. 

\subsection{Source Coding}
Given a quantized representation $\mathbf{z}_Q$, the discrete probabilities $p_i$ for all codebook symbols are estimated from feature representations of a fixed, finite set of training channels, and \textit{lossless} source coding is applied during inference for storage or relaying purposes. Our method is compatible with any source coding scheme. In practice, we use arithmetic coding \cite{mentzer2019practical} due to its near-optimal performance and extremely efficient publicly available implementation \cite{aenc_github}. Since the coding is lossless, there is no incurred performance loss.

\subsection{Decoder}
The decoder $g$ is a deep neural network with an architecture that mirrors $f$, including the number of layers and the hidden dimension. That is, it maps an input latent matrix $\mathbf{z}$ to the reconstructed soft bit matrix $\tilde{\mathbf{\Lambda}}$ by applying the shared layers in a row-wise manner:
\begin{equation}
    \tilde{\mathbf{\Lambda}} = \textrm{stack}_i(g(\mathbf{z_i})).
\end{equation}

Given all components, the model is trained with the end-to-end supervised loss:
\begin{equation}
    L(\mathbf{\Lambda}, \tilde{\mathbf{\Lambda}}; \tau) = \mathcal{D}(\mathbf{\Lambda}, \tilde{\mathbf{\Lambda}}) + \alpha \mathcal{H}(\mathbf{z}; \tau).
\label{eq:loss_fn}
\end{equation}
The first term corresponds to the quantization-aware reconstruction loss that ensures soft bits are recovered properly after numerical quantization of the latent representation $\mathbf{z}$. The second term serves as a approximation for minimizing the entropy of the quantized latent representation $\mathbf{z}_Q$ and to enable further gains with source coding, where $\alpha$ is a hyper-parameter that directly controls the rate-distortion trade-off.

\section{Experimental Results}
\subsection{Architecture and Training}
We use deep neural networks for $f$ and $g$, each with four hidden layers, $\mathrm{relu}$ hidden activations, $\tanh$ activation at the output (for both $f$ and $g$), and a hidden size of $4K$, where $K$ is the modulation order for which we train the method -- as well as the input size to the network. The latent dimension is always three and we initialize all layers with the Glorot scheme \cite{glorot2010understanding} to match the conditions of Theorem~\ref{theorem:main}. Complete details about the architecture are found in our code repository linked in the abstract.

The latent quantizer $\mathcal{Q}$ uniformly covers the interval $[-0.8, 0.8]$ using a number of $64$ codebook entries ($6$ bits), and remains fixed throughout the entire training and inference procedures. The same $\mathcal{Q}$ is used for all the latent dimensions and the choice of the interval is a direct consequence of the range of the latent representation under the $\tanh$ operator, as shown in Fig.~\ref{fig:theory}. It can be seen that the latent code is bound to this interval, hence no codebook collapse occurs at initialization.

The data used to train all models comes from transmissions across i.i.d. Rayleigh fading channels, where $h_i \sim \mathcal{N}_\mathbb{C}(0, 1)$ and the noise $n_i \sim \mathcal{N}_\mathbb{C}(0, \sigma_n)$. Payloads are generated by randomly sampling bits with equal probabilities and codewords are obtained by using a low-density parity check (LDPC) code of size $(324, 648)$, for a total of $100000$ training codewords at uniformly spaced SNR values. Importantly, our method is \textit{only} trained on soft bits from i.i.d. channels, and is not trained on a specific wideband channel distribution. We find that a range of $\alpha$ between $0.001$ and $0.03$ generally covers the entire rate-distortion curve, and we anneal $\tau$ at epoch $t$ by the schedule $\tau_t = 40 \times 1.001^t$.

A single network is trained across the entire SNR range, and takes about three hours for $2000$ epochs (invariant to $K$) on an NVIDIA RTX 2080Ti GPU. Inference takes less than $1$ ms for an OFDM wideband channel with $108$ subcarriers. Storing the network for $K=6$ takes a total of $82.8$ kB in floating point precision. During inference, soft bits are quantized and reconstructed, and belief propagation decoding is performed to obtain a complete communication chain. We measure end-to-end performance through the block (codeword) error rate figure.

We train the baseline in \cite{arvinte2019deep} by using exactly the same data and backbone architecture for a fair comparison. We also compare with the optimal scalar method in \cite{winkelbauer2015quantization} by learning a separate quantization codebook for each soft bit position, at each SNR value. This partially compensates for the extra learnable parameters that deep learning methods have.

\subsection{End-to-End Quantization Performance}
\begin{figure}[!t]
\centering
\includegraphics[width=0.93\linewidth]{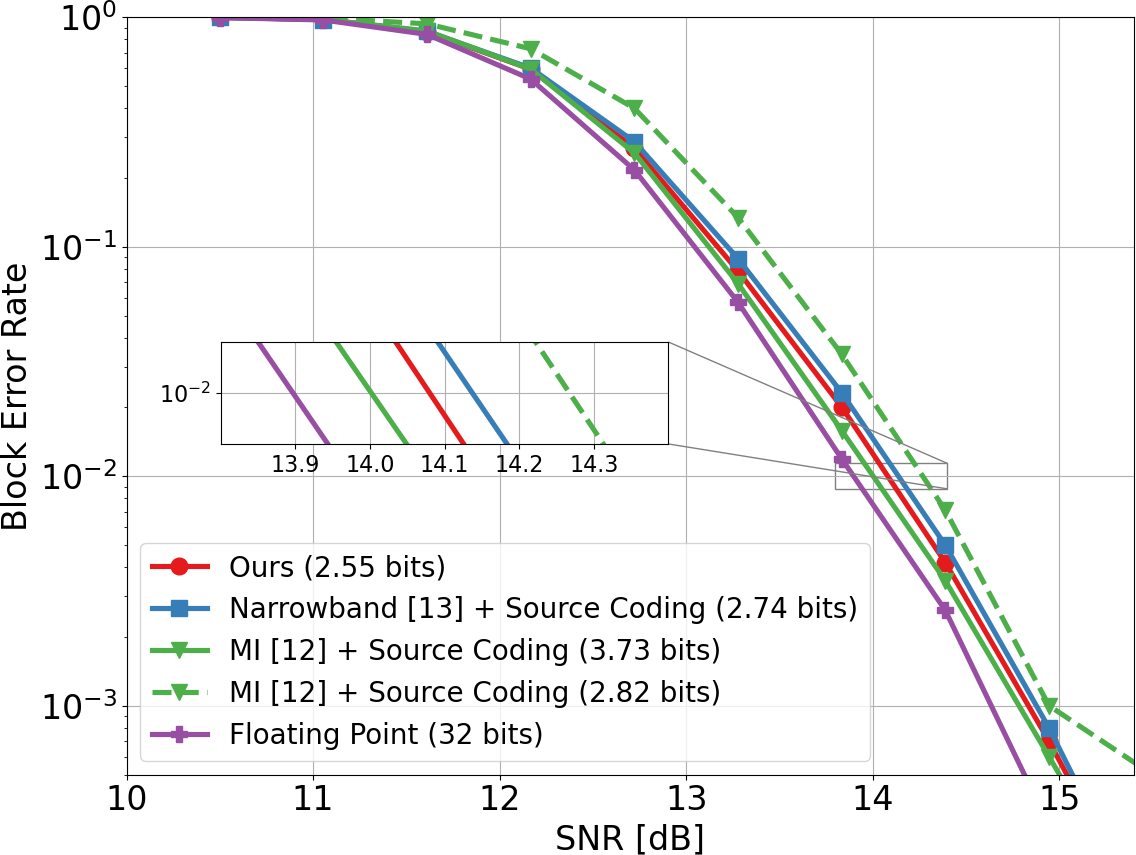}
\caption{Block error rate as a function of SNR for the proposed method, baselines and floating point (no quantization), for $K=6$ ($64$-QAM) modulation in EPA channels with a bandwidth of $10$ MHz. The values in parentheses indicate the average storage cost per soft bit.}
\label{fig:op_curve}
\end{figure}

Fig.~\ref{fig:op_curve} shows the performance of all methods in a $K=6$ (64-QAM) modulation scheme and EPA wideband channel model with $108$ subcarriers allocated per codeword, at a carrier frequency of $2$ GHz and channel bandwidth of $10$ MHz. The number in the parentheses indicates the average cost required to store a single soft bit, where we average this cost over the range of SNR values that lead to block error rates between $1$ and $0.001$, since this range is of practical interest. A key takeaway here is that all methods are calibrated to produce the same end-to-end performance, with minimal deviations from un-quantized performance. The proposed approach suffers a performance loss of $0.18$ dB compared to floating point at a target error rate of $0.01$, and has the same performance as \cite{arvinte2019deep}, while achieving an average compression gain of $7\%$. Both deep learning-based methods greatly surpass the scalar quantizer, with ours having an average compression gain of $31\%$.

\begin{figure}[!t]
\centering
\includegraphics[width=0.93\linewidth]{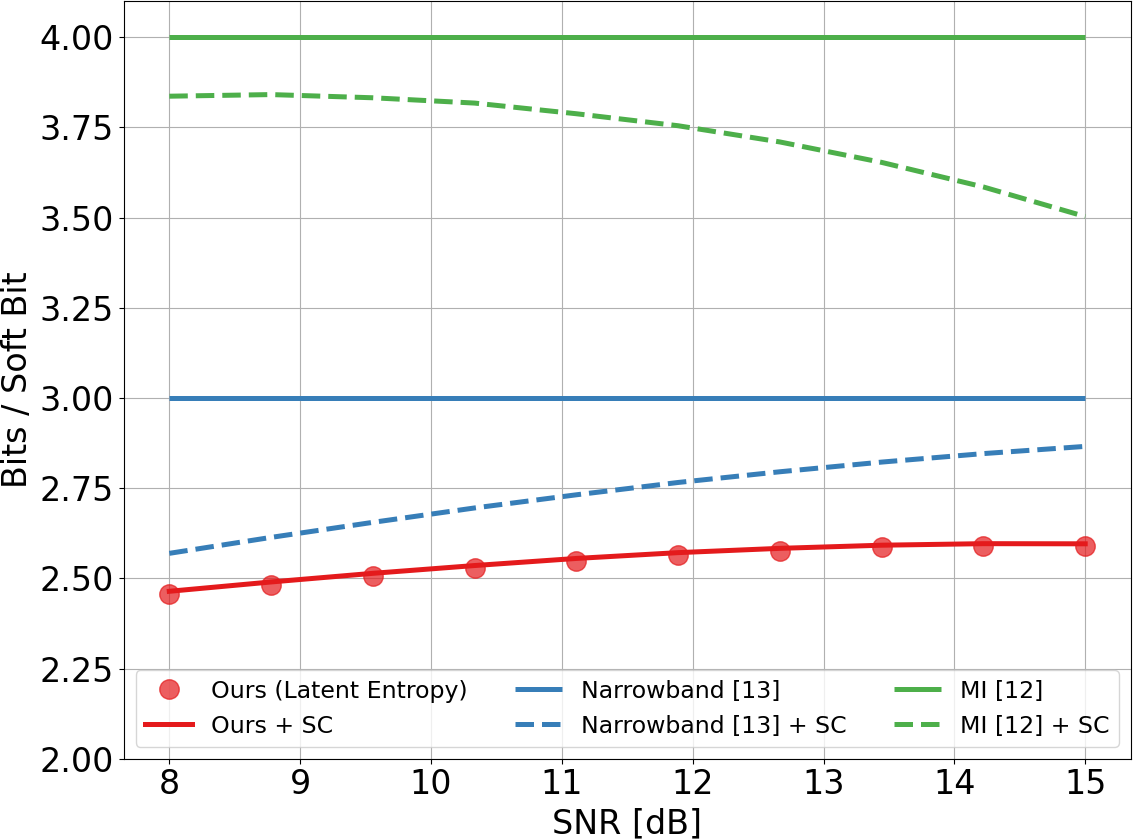}
\caption{Soft bit quantization cost as a function of operating SNR for the proposed method and the baselines in EPA channels. The red and black solid lines indicate baselines without source coding, while the dashed lines are with source coding. The orange curve is the performance of our method, while the blue dots indicate the estimated entropy of $\mathbf{z}_Q$ (lowest possible rate). For all methods, no specific EPA channel simulations are used to train the models or calibrate the source coding.}
\label{fig:q_perf}
\end{figure}

Fig.~\ref{fig:q_perf} reveals how quantization cost scales with SNR for the different methods, as well as the near-optimality of arithmetic coding in a wideband scenario. For scalar quantization methods such as maximum MI \cite{winkelbauer2015quantization}, the cost per soft bit decreases with increasing SNR. Asymptotically, this behaviour is optimal since as $\textrm{SNR} \rightarrow \infty$, then the soft bits become discrete binary random variables as in Theorem~\ref{theorem:main}, and one bit per soft bit is the optimal quantization scheme. This trend is opposite for deep learning methods, since the same model accommodates the entire SNR regime: there, the average cost per bit \textit{increases} as the SNR increases, and this phenomenon is much more pronounced for the baseline in \cite{arvinte2019deep}. We find that the proposed approach helps counteract this sub-optimality, again due to its entropy objective in the loss function.

Fig.~\ref{fig:q_perf} also plots the performance of the two baselines with and without (horizontal lines) source coding. We note that, while source coding benefits both baselines, the proposed approach still improves compression rates in a broad SNR range due to the entropy-aware nature of (\ref{eq:loss_fn}). In the high SNR regime, the proposed method achieves compression gains of up to $10\%$ compared to \cite{arvinte2019deep}, and the source coding is near-optimal, since it achieves the entropy marked with circles.

\subsection{Rate-Distortion Trade-Off}
\begin{figure}[!t]
\centering
\includegraphics[width=0.93\linewidth]{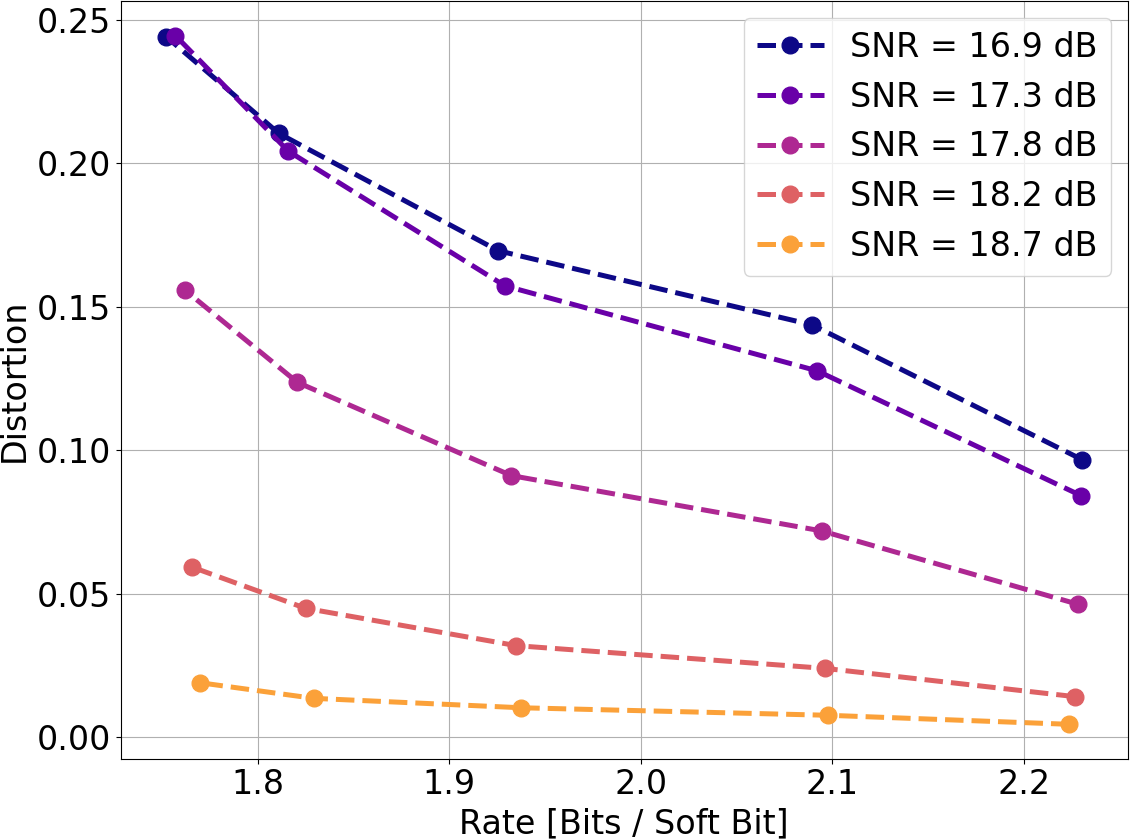}
\caption{Rate-distortion curves obtained by tuning $\alpha$ in the loss function, for $K=8$ ($256$-QAM) in EPA channels. In this figure, $\alpha$ is linearly interpolated between $0.01$ (right-most points) and $0.03$ (left-most points) with a spacing of $0.005$. The y-axis represents the \textit{additive} block error rate incurred against a floating point solution. Each of the curves represents a specific SNR point.}
\label{fig:rd_curves}
\end{figure}

Fig.~\ref{fig:rd_curves} investigates the impact of $\alpha$ during training our method, for $K=8$ on EPA channels with $81$ subcarriers (since the bit mapping is denser than $K=6$, fewer channel uses are required to send a packet). The operating characteristics of the model are close to the ones in Fig.~\ref{fig:op_curve}, and extended results can be found in our code repository.

To control the trade-off between rate and distortion, we vary the $\alpha$ parameter in our loss function between the range of $0.01$ and $0.03$ and train a separate model at each value. From Fig.~\ref{fig:rd_curves} it can be noticed that, at lower SNR values the absolute penalty in end-to-end error is larger if we quantize aggressively, whereas the error increases are much smaller in the high SNR regime.

\section{Conclusion}
In this paper, we have introduced a deep learning approach for wideband soft bit quantization. Our formulation included a fixed quantizer and a quantization- and entropy-aware training objective, as well as the use of source coding at inference. Our theoretical results proved that a fixed quantizer is sufficient for efficient training, and the experiments have shown state-of-the-art quantization performance in a wide SNR range, as well as flexibility in controlling the rate-distortion trade-off.

The model is compact and inference is efficient, achieving sub-ms latency for an entire wideband channel. Our model is also not trained on a specific channel distribution, but instead can operate on arbitrary wideband channels. While this provides a degree of flexibility, a promising future research direction is to investigate whether further compression gains can be obtained by specializing a model for a \textit{specific} channel distribution and develop adaptive quantization schemes.

\bibliographystyle{IEEEtran}
\bibliography{myBib}

\end{document}